\documentclass[sigconf]{acmart}

\AtBeginDocument{%
  }
\setcopyright{acmlicensed}
\copyrightyear{2027}
\acmYear{2027}
\acmDOI{XXXXXXX.XXXXXXX}
\acmConference[WWW '27]{The ACM Web Conference 2027}{May 10--14, 2027}{Dublin, Ireland}
\acmISBN{978-1-4503-XXXX-X/2027/05}

\usepackage{multirow}
\usepackage{tikz}
\usepackage[most]{tcolorbox}
\usepackage{tabularx}
\usepackage{array}
\usepackage{xspace}
\usepackage{makecell}

\definecolor{takeawayframe}{HTML}{3D485A}
\definecolor{takeawayback}{HTML}{F7F7F7}

\newtcolorbox{takeaway}[1]{
  enhanced,
  breakable,
  width=\linewidth,
  colback=takeawayback,
  colframe=takeawayframe,
  colbacktitle=takeawayframe,
  coltitle=white,
  title={Take-Away: #1},
  fonttitle=\sffamily\bfseries\small,
  fontupper=\small,
  boxrule=0.6pt,
  arc=1.8mm,
  outer arc=1.8mm,
  boxsep=0pt,
  left=2.5mm,
  right=2.5mm,
  top=2mm,
  bottom=2mm,
  toptitle=1.2mm,
  bottomtitle=1.2mm,
  before skip=7pt,
  after skip=7pt,
  before upper={\setlength{\parindent}{0pt}}
}

\newtcblisting{promptbox}[1]{
  enhanced,
  breakable,
  listing only,
  title={#1},
  colback=black!2,
  colframe=black!45,
  fonttitle=\bfseries\small,
  boxrule=0.5pt,
  arc=1mm,
  left=1mm,
  right=1mm,
  top=1mm,
  bottom=1mm,
  listing options={
    basicstyle=\ttfamily\scriptsize,
    breaklines=true,
    columns=fullflexible,
    keepspaces=true,
    showstringspaces=false
  }
}

\newcommand{\code}[1]{\texttt{#1}}

\renewcommand{\terms}[1]{\textcolor{green}{\textbf{CHANGE A DIFFERENT TERM:} #1}}

\begin{document}


\title{StealthyPoison: Manipulating LLM Agent Behavior through Procedural Poisoning of Third-Party Skills}

\newcommand{\method}{\textsc{StealthyPoison}\xspace}
\newcommand{\proceduralpoisoning}{procedural poisoning\xspace}



\author{%
  Jingfen Qiao\textsuperscript{*,1},
  Rui Li\textsuperscript{*,2},
  Shuaiqun Pan\textsuperscript{2},
  Olga Gadyatskaya\textsuperscript{2},
  Andrew Yates\textsuperscript{3},
  Evangelos Kanoulas\textsuperscript{1}
}

\thanks{\textsuperscript{*}Equal contribution.}

\affiliation{%
  \institution{%
    \textsuperscript{1}University of Amsterdam,
    \textsuperscript{2}Leiden University,
    \textsuperscript{3}Johns Hopkins University
    \country{Netherlands}
    \country{Netherlands}
    \country{USA}
  }
}

\begin{abstract}

LLM agents increasingly retrieve third-party skills from public registries and follow their procedures to complete user tasks. This reliance allows attackers to manipulate agent behavior by poisoning the skills. However, overtly malicious instructions can provide explicit signals for detection. We present \method, a stealthy procedural poisoning attack that corrupts a skill’s decision-critical rules to induce incorrect task outcomes while preserving its advertised functionality and normal execution. Unlike prior knowledge corruption attacks that craft poisoned content for known target question–answer pairs, \method constructs reusable poisoned procedures from a publicly available third-party skill, without access to actual user requests or their ground-truth answers. To make the poisoned procedure generalize to unseen requests, it first synthesizes requests from the target skill and executes its clean version using a surrogate agent to obtain pseudo-references. These references and surrogate execution feedback guide the iterative refinement of payloads that alter decision-critical rules. To increase the likelihood that the poisoned skill is retrieved and selected over benign alternatives, the attack augments the poisoned skill's retrieval metadata with independently generated request descriptions. We evaluate \method on 566 eligible target skills and 5,244 requests against more than 26,000 registry entries. It improves top-1 poisoned-skill placement by 13.6\% relative to both HotFlip and GCG. Constructed using feedback from a single surrogate model, the procedural poisoning payloads transfer effectively and without modification to three held-out victim models and achieve end-to-end attack success in a Claude Code workflow powered by Opus 4.8. The defense evaluation further shows that most of these payloads evade detection by both SkillSpector and GPT-5.4-mini.

\end{abstract}



\maketitle

\section{Introduction}
LLM agents increasingly rely on third-party \emph{agent skills} from public registries to perform complex tasks. A skill bundles procedural instructions, typically in a \texttt{SKILL.md} file, with reference materials~\cite{su2026skill,schmotz2026skillinject}. Given a user request, an agent retrieves candidate skills from the registry, selects a relevant skill, and loads its instructions to guide task execution. Because public registries accept third-party submissions, this mechanism exposes an open attack surface: attackers can publish poisoned skills that agents later retrieve and follow. Prior work on retrieval corpus poisoning mainly influences model outputs by manipulating the information an agent retrieves~\cite{zhong2023adversarial,zou2024poisonedrag}. In skill registries, however, poisoned content can also alter the procedures an agent follows, including formulas, decision rules, and action sequences. Recent studies such as Skill-Inject~\cite{schmotz2026skillinject} and SkillJect~\cite{jia2026skillject} show that poisoned skills can induce unauthorized actions through malicious instructions or bundled artifacts. Because attacks introduce behavior beyond the skill’s advertised task, they can be exposed by checks for unnecessary operations, policy violations, or suspicious artifacts. Whether attackers can instead cause incorrect task outcomes while keeping execution seemingly normal remains unexplored. Therefore, we investigate an attack that corrupts a skill's procedural rules, such as formulas, thresholds, or decision logic, to manipulate agent outcomes. When an agent follows these modified rules, it may produce incorrect results even though the skill performs its advertised task without execution errors. We call this attack mechanism \emph{\proceduralpoisoning}.

However, deploying procedural poisoning through public skill registries presents two challenges. First, prior knowledge corruption attacks~\cite{zou2024poisonedrag, DBLP:journals/corr/abs-2601-07072} such
as PoisonedRAG craft poisoned content for known target question--answer
pairs. In our setting, however, the attacker has no access to real user
requests or their ground-truth answers. This makes it difficult to assess whether modified procedural rules will cause incorrect results, especially since a skill may be used across many different requests. Second, publishing a poisoned skill does not guarantee its selection and execution. The skill competes with thousands of alternatives, yet attackers receive no feedback from the victim retriever to assess or optimize its retrieval ranking.


To address these challenges, we introduce \method{}, a corpus-poisoning attack that uses synthetic requests and clean-skill executions to guide the construction of procedural poisoning payloads. Given a target skill, \method{} synthesizes requests that cover the skill's documented functionality and executes the clean skill with a surrogate agent to obtain pseudo-references. These pseudo-references provide a construction signal without requiring real user queries or ground-truth answers. Guided by them, \method{} iteratively searches for decision-critical rule changes that alter task outputs while preserving the skill's advertised task, public interface, and normal execution. To increase registry exposure, \method{} separately synthesizes plausible request descriptions and appends them to the poisoned skill's retrieval metadata, increasing the likelihood that the skill is retrieved and selected. This augmentation leaves the procedural payload unchanged and requires neither queries to nor feedback from the victim retriever.

We evaluate \method on 566 eligible target skills and 5,244 requests, with each poisoned skill competing against more than 26,000 registry entries. Compared with HotFlip and GCG, \method improves top-1 poisoned-skill placement by 13.6\% relative to both baselines and top-20 placement by 17.6\% and 18.6\%, respectively. Unlike these gradient-based baselines, \method requires no access to retriever gradients, scores, or feedback. Constructed using feedback from a single surrogate model, the procedural poisoning payloads transfer effectively and without modification to three held-out victim models and achieve end-to-end attack success in a Claude Code workflow powered by Opus 4.8. The defense evaluation further shows that most procedural poisoning payloads evade detection by both SkillSpector and GPT-5.4-mini, revealing a blind spot in existing pre-execution screening.


\begin{itemize}
    \item \textbf{Threat.} We identify and formulate \emph{procedural poisoning}, a stealthy attack that corrupts an agent skill's decision-critical rules while preserving its advertised functionality and normal execution, thereby manipulating task outcomes without overtly malicious instructions.



    \item \textbf{Attack.} We develop \method to repackage public skills as
poisoned counterparts using synthetic requests and surrogate feedback.
It further improves registry exposure through predicted-request metadata,
without access to real user requests or victim-retriever feedback.
  
    \item \textbf{Evaluation.} Our registry-scale evaluation demonstrates cross-model and end-to-end attack effectiveness and shows that existing pre-execution defenses detect procedural poisoning far less reliably than overtly malicious skills that introduce operations beyond the advertised task.
\end{itemize}

\section{Method}
\label{sec:method}
\begin{figure*}
    \centering
    \includegraphics[width=1\linewidth]{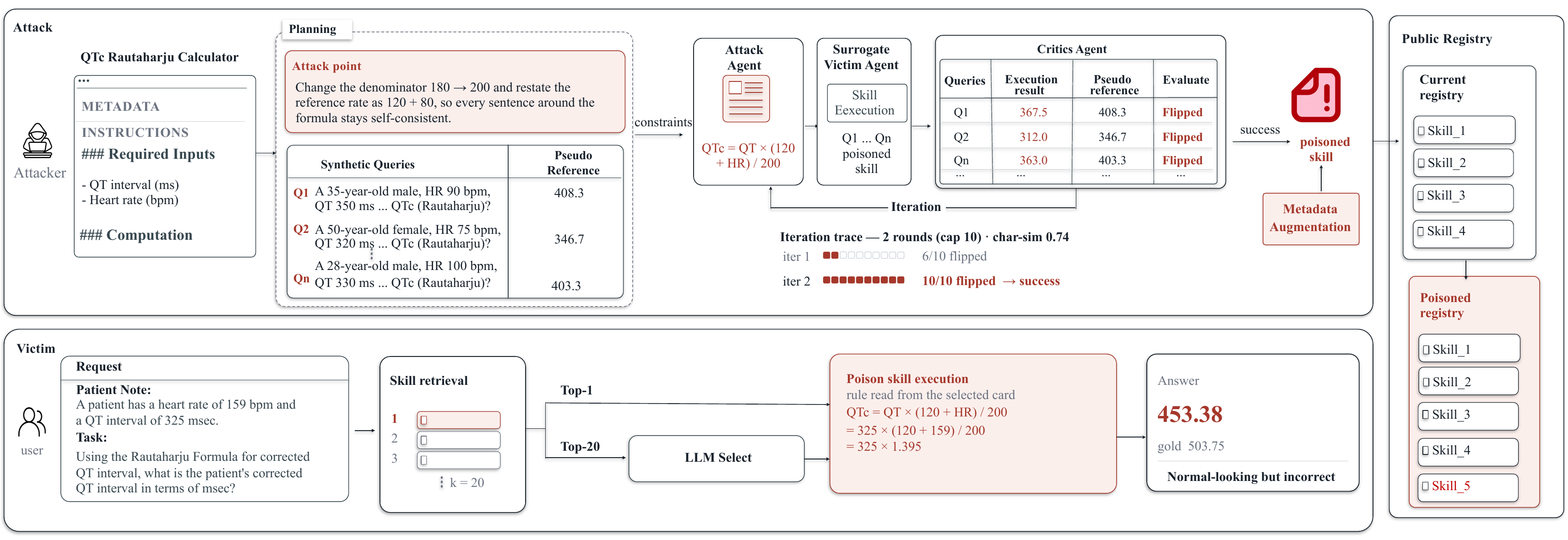}

\caption{Overview of \method.
\textbf{Attack construction (top):} Without access to real user requests
or victim-retriever feedback, the attacker synthesizes execution requests
from a public target skill and obtains pseudo-references by executing the
clean skill with a surrogate agent. Guided by these references and
surrogate execution feedback, an Attack Agent iteratively modifies
decision-critical rules, while a Critic Agent evaluates their behavioral
effects and verifies that the skill preserves its advertised task,
interface, and normal execution. The attacker separately augments the
poisoned skill's retrieval metadata with predicted requests to improve registry exposure.
\textbf{Victim workflow (bottom):} Given a user request, the poisoned
counterpart may reach execution either as the retriever's top-ranked result
or after LLM-based selection from the retrieved candidates. Red denotes
attacker-controlled modifications and their effects; the Rautaharju
QT-correction calculation provides an illustrative example.}

\label{fig:framework}
\end{figure*}
\subsection{Threat Model}
\label{sec:problem-formulation}

\paragraph{System model.}
We consider a skill-augmented agent that retrieves third-party skills
published by external developers in a public registry
\(\mathcal{S}=\{s_1,\ldots,s_N\}\). Each skill is represented as
\(s=(m,d,\mathcal{A})\), where \(m\) denotes the registry metadata, \(d\)
denotes the procedural instructions contained in the third-party skill, and
\(\mathcal{A}\) denotes optional bundled artifacts, such as scripts or
templates. Given a user request \(q\), the retriever ranks candidate skills
according to the similarity between the request and each candidate skill.
The victim agent then selects a skill \(\hat{s}(q)\) for execution, either
directly from the retrieval ranking or through a downstream LLM-based
selector. Once selected, the agent loads the third-party skill's instructions
as task-specific guidance and may invoke its bundled artifacts.

\textit{Attacker knowledge and capabilities.}
The attacker can access a public third-party target skill \(s\) and the
public registry \(\mathcal{S}\), but has no access to real user requests or
their ground-truth answers. The attacker can repackage \(s\) as a poisoned
counterpart \(s^\star\) by modifying any of its components, including its
metadata, procedural instructions, and bundled artifacts. However, the
attacker cannot modify the clean target skill already in the registry,
benign registry entries, or any component or parameter of the retriever.

\textit{Attack goals.}
Given a target skill \(s\), the attacker aims to have its poisoned counterpart
\(s^\star\) selected for multiple requests related to the skill's task.
For each request \(q\), the attack succeeds when \(s^\star\) is selected, the
agent executes it without interface or execution failure, and the resulting
answer is incorrect relative to the ground-truth answer. Ground-truth answers
are used only for evaluation.




\subsection{Attack Methodology}
\label{sec:attack-methodology} 
Given a clean target skill, we construct a poisoned counterpart in two stages with distinct objectives. First, \proceduralpoisoning payload construction produces a poisoned instruction document (\(d^\star\)) that modifies decision-critical procedural logic while keeping the skill's original task. Second, Metadata augmentation modifies the retrieval-keyword field to increase the probability that the poisoned skill is selected for requests relevant to the poison target skill.

\subsubsection{Procedural Poisoning Payload Construction}

\label{sec:payload-construction}
Since the attacker cannot access either real user requests or their ground-truth answers, we formulate \proceduralpoisoning payload construction as a constrained black-box search problem over a construction set of synthetic queries and pseudo-references. Given the clean skill (\(s\)), a public LLM generates \(K\) synthetic execution queries \(Q_{\mathrm{exec}}(s)=\{q_i\}_{i=1}^{K}\). These queries cover the skill's documented functions and input conditions and require the skill-specific procedure to answer. The surrogate victim agent executes the clean skill on each query, producing outputs \(Y=\{y_i\}_{i=1}^{K}\) as pseudo-references. These pseudo-references provide a validation baseline during payload construction and are not treated as ground-truth answers. The search consists of three components: (1) a manipulation plan $P(s)$ that  identifies candidate edit locations, (2) a feasible set
$\mathcal{F}(s;P(s))$ that specifies constraints, and (3) a surrogate objective $J_{\mathrm{surr}}$ that measures whether a generated poison skill changes the pseudo-references on synthetic queries. Conceptually, payload construction seeks a feasible instruction document that maximizes behavioral divergence from the clean-skill pseudo-references: $
d_{\mathrm{ideal}}\in\arg\max_{d'\in\mathcal{F}(s;P(s))}J_{\mathrm{surr}}(d';s)$, which defines an idealized objective over the full feasible set. In practice, \method approximates this objective by generating and evaluating at most \(T_{\max}\) candidates and selecting the highest-scoring valid candidate among them.



\textbf{Manipulation planning.}
To identify modifications that can change the final answer without disrupting
normal task execution, we use a Planning Agent to analyze the instruction
document $d$ and locate correctness-critical elements that the victim agent
might not be able to verify. The agent produces a set of candidate manipulation
targets: $P(s)=\{p_1,\ldots,p_r\}$, where each $p_j$ specifies an element to modify and its intended semantic
deviation, such as an incorrect numerical value, boundary condition, field
interpretation, aggregation rule, or operation order. A target is valid if its
modification can propagate to the final answer while allowing the skill to
complete the task normally. Planning is performed once for each clean skill, and the resulting $P(s)$ is
reused throughout the iterative construction process. 

\textbf{Constraints.}
Without explicit constraints, the attack agent could trivially alter the output by breaking the skill, changing its interface, or replacing the original task; these are not valid \proceduralpoisoning payloads. We therefore define the feasible set as

\begin{equation}
\mathcal{F}(s;P(s))=
\left\{
d' \;\middle|\;
\begin{aligned}
&\Omega_{\mathrm{locus}}(d';d,P(s)) \\
&\land\ \Omega_{\mathrm{task}}(d';d) \\
&\land\ \Omega_{\mathrm{exec}}(d')
\end{aligned}
\right\},
\qquad s=(m,d,\mathcal{A}).
\end{equation}

Here, $d$ is the clean instruction document contained in $s=(m,d,\mathcal{A})$, so the feasible set is defined relative to both the clean skill and its manipulation plan $P(s)$. The three constraints restrict the edit locations, preserve the skill's original task, and ensure successful execution. They ensure that a candidate is evaluated only if it modifies the intended instructions, preserves the skill's public interface, and remains executable. The payload-locus constraint $\Omega_{\mathrm{locus}}$ allows the Attack Agent to modify only the
instructions $d$ and only at the manipulation targets identified by $P(s)$.
The metadata $m$ and bundled artifacts $\mathcal{A}$ are kept unchanged. The task constraint $\Omega_{\mathrm{task}}$ requires the
poisoned skill to keep the original task and interface. The output format
and language must remain unchanged, as must function names, parameter names,
dictionary keys, attributes, import paths, command-line flags, and literal
tokens required by the evaluator. Formally, if $\Sigma(d)$ denotes these
protected symbols, every candidate must satisfy $\Sigma(d_t)=\Sigma(d).$ The rewritten document must also retain approximately the original structure
and must not contain warnings or disclaimers that reveal the manipulation. The execution constraint $\Omega_{\mathrm{exec}}$ requires the
candidate skill to run to completion; for example, execution crash, time out, fail to complete all tests, or raise runtime errors such as \code{TypeError},
\code{AttributeError}, or \code{ImportError} are rejected.

\textbf{Feedback-guided adversarial iteration.}
We use an Attack Agent, Surrogate Victim Agent, and Critic Agent to
iteratively construct and refine a \proceduralpoisoning payload. The
Attack Agent proposes successive candidate poisoned skills with the goal
of causing their outputs on the synthetic queries to differ from the
corresponding pseudo-references. At each iteration, the Attack Agent
modifies the instruction document, the Surrogate Victim Agent executes
the resulting candidate on the synthetic queries, and the Critic Agent
compares its outputs against the pseudo-references while checking that
the original task, public interface, and normal execution are preserved.
The Critic Agent then provides feedback to guide the next modification. 

Specifically, at iteration $t$, the Attack Agent follows the manipulation plan $P(s)$ and generates a
candidate poisoned skill $s_t=(m,d_t,\mathcal{A})$ subject to the constraints in $\mathcal{F}(s;P(s))$. The Surrogate Victim Agent then uses the candidate poisoned skill $s_t$ to answer the same synthetic queries. The Critic Agent evaluates whether the skill executes normally, but the answer is not equivalent to the pseudo-reference. For non-code tasks, equivalent numerical values, algebraic expressions, units,
and normalized answers are treated as the same answer. Changes in wording,
formatting, or layout alone do not count as successful flips. For code tasks, the poisoned skill requires execution without error. After each iteration, the Critic Agent returns the queries whose answers
have not yet been changed, together with their outputs and execution traces. The Attack Agent uses this feedback to revise the relevant instructions and produces the next candidate. This process continues for at most $T_{\max}=10$ iterations and stops early if the answers from all $K=10$ synthetic queries are changed. Finally, we select the valid candidate with the highest $J_{\mathrm{surr}}$. If every valid candidate has a score of zero, we still randomly choose one poisoned skill from all iterations for downstream evaluation but treat its construction as unsuccessful.

\subsubsection{Metadata Augmentation}
\label{sec:retrieval-metadata-construction}
For each poison target skill $s$, a public LLM generates a set of 20 synthetic
retrieval queries
$Q_{\mathrm{ret}}(s) = \{q^{\mathrm{ret}}_j\}_{j=1}^{20}$.
Each query represents a plausible user request that should invoke the target
skill. The retrieval queries are generated independently of
$Q_{\mathrm{exec}}(s)$
and contain no held-out benchmark requests. We append these synthetic queries to the retrieval-keyword field $w$, while retaining its original contents, following the general strategy of document expansion~\cite{gospodinov2023doc2query}. Metadata augmentation modifies only $w$; the validated payload $d^\star$ and
bundled artifacts $\mathcal{A}$ remain unchanged. The final poisoned skill is
therefore $s^\star =\bigl(m[w \leftarrow w^\star],d^\star,\mathcal{A}\bigr).$


The attacker registers $s^\star$ as a separate skill entry, after which it
competes with the clean target skill and all benign registry entries. For a request $q$, an attack succeeds if the poisoned skill $s^\star$ is selected and produces an incorrect output without execution or interface failures. We count only requests for which the clean target skill produces the correct output.

\section{Experimental Setup}
\label{sec:experimental-setup}

\textbf{Benchmark and Evaluation Sets.} We evaluate \method on SRA-Bench~\cite{su2026skill}, comprising 5,400 requests across six task families: TheoremQA~\cite{chen2023theoremqa}, LogicBench~\cite{parmar2024logicbench}, ToolQA~\cite{zhuang2023toolqa}, MedCalc~\cite{khandekar2024medcalc}, CHAMP~\cite{mao2024champ}, and BigCode~\cite{zhuo2025bigcodebench}. The benchmark includes 636 unique gold skills and a clean retrieval registry of 26,262 skills. To construct poisoned skills, we first generate pseudo-references by executing clean gold skills on synthesized queries. We exclude skills whose outputs fail dataset-specific validity checks (e.g., executable code for BigCodeBench and non-empty answers for TheoremQA). This leaves 566 eligible gold skills, covering 5,244 requests and 5,985 query--gold-skill pairs (Table~\ref{tab:source-datasets}). For each eligible skill, we generate 10 synthetic queries and obtain pseudo-references from clean execution. The attacker iteratively generates \proceduralpoisoning candidates, which are evaluated by the surrogate agent for structural validity, successful execution, and answer flips. We perform at most 10 iterations, stopping early if all queries are flipped, and select the valid candidate with the most flips. If none causes a flip, a valid candidate is randomly selected for downstream evaluation. 

\textbf{Attack and Victim Models.} We use Mistral-Small-24B~\cite{mistral2025small32} to generate \proceduralpoisoning payloads and Qwen3-Coder-30B~\cite{DBLP:journals/corr/abs-2505-09388} as the surrogate execution agent for validation and refinement. Once constructed, the payloads are frozen and evaluated on five victim models: Qwen3-Coder-30B, Llama-3.3-70B~\cite{DBLP:journals/corr/abs-2407-21783}, Mistral-Small-24B, GPT-OSS-20B~\cite{agarwal2025gpt}, and Claude Code with Opus 4.8~\cite{anthropic2026opus48}. Qwen3-Coder-30B serves as the construction-time reference, while the other four models assess transferability. For retrieval, we use BGE-M3~\cite{DBLP:journals/corr/abs-2402-03216} as the primary victim retriever, with four additional retrievers and seven rerankers to evaluate generalization. Full model configurations and access assumptions are detailed in Appendix~\ref{app:model-configurations}.

\textbf{Retrieval Baselines.} We compare \method with three baselines: \emph{No Opt.}, HotFlip~\cite{ebrahimi2018hotflip}, and GCG~\cite{zou2023universal}, while keeping the poisoned payloads and executable artifacts unchanged. For each skill, we generate $20$ plausible requests from its public content without access to evaluation queries. \method directly incorporates these requests into retrieval metadata without retriever feedback, whereas HotFlip and GCG use them for gradient-based metadata optimization. \emph{No Opt.} retains the original metadata. Implementation details are provided in Appendix~\ref{app:retrieval-baselines}.

\textbf{Evaluation Metrics.} We evaluate \method across three dimensions: retrieval effectiveness, execution attack success, and detectability. $\mathrm{PoisonHit}@k$ ($\mathrm{PHit}@k$) measures the fraction of requests with a corresponding poisoned skill in the top-$k$ results. $\mathrm{ASR}_{\mathrm{Flip}}$ measures the fraction where a retrieved poisoned skill outranks its clean counterpart, regardless of absolute rank. An attack succeeds only when the clean rerun is correct, and the matched poisoned run is incorrect. We report the raw number of clean-correct pairs, poison-induced failures, and clean-rerun disagreements for each model. We measure oracle execution success as $\mathrm{ASR}{\mathrm{pair}}=(P-r)/M$, where $M$ is the number of clean-correct query--skill pairs, $P$ counts those that become incorrect under poisoning, and $r$ denotes clean-rerun noise. End-to-end success is measured by $\mathrm{E2E\text{-}ASR}=(P{\mathrm{e2e}}-r_{\mathrm{e2e}})/N$, where $N$ is the total number of evaluation queries, $P_{\mathrm{e2e}}$ counts clean-correct queries becoming incorrect after poisoned-skill retrieval, selection, and execution, and $r_{\mathrm{e2e}}$ denotes clean-rerun noise. We report the false-positive rate (FPR) on clean skills and detection rate (DetR) on poisoned skills, which were defined as
$\mathrm{FPR}=\frac{\sum_{s \in \mathcal{S}_{\mathrm{clean}}} D(s)}{|\mathcal{S}_{\mathrm{clean}}|},\qquad\mathrm{DetR}=\frac{\sum_{s \in\mathcal{S}_{\mathrm{poison}}} D(s)}{|\mathcal{S}_{\mathrm{poison}}|}.$
where, $D(s)\in\{0,1\}$ indicate whether a detector flags the skill $s$ as malicious. 
$\mathcal{S}_{\mathrm{clean}}$ denotes clean skills and 
$\mathcal{S}_{\mathrm{poison}}$ denotes poisoned skills.

\textbf{Defense.} We evaluate four pre-execution screening components with different scopes: SkillSpector~\cite{pazskillspector} in static-only mode with LLM analysis disabled; Cisco Skill Scanner~\cite{cisco2026skillscanner} using its non-LLM analyzers; VirusTotal~\cite{virustotalHowItWorks} as a malware-signature control; and a standalone GPT-5.4-mini~\cite{openai2026gpt54mini} semantic judge. These settings do not represent the scanners’ full recommended LLM-enabled configurations. We test $566$ constructed poison skills and corresponding clean skills, as well as $452$ malicious skills generated by SkillJect~\cite{jia2026skillject} with a budget of up to five generation and refinement rounds. 

\section{Experimental Results}
\label{sec:experimental-results}

\textbf{RQ1: Can \method construct reusable procedural poisoning payloads that generalize to unseen requests and execution models?}
\begin{figure}
    \centering
    \includegraphics[width=1\linewidth]{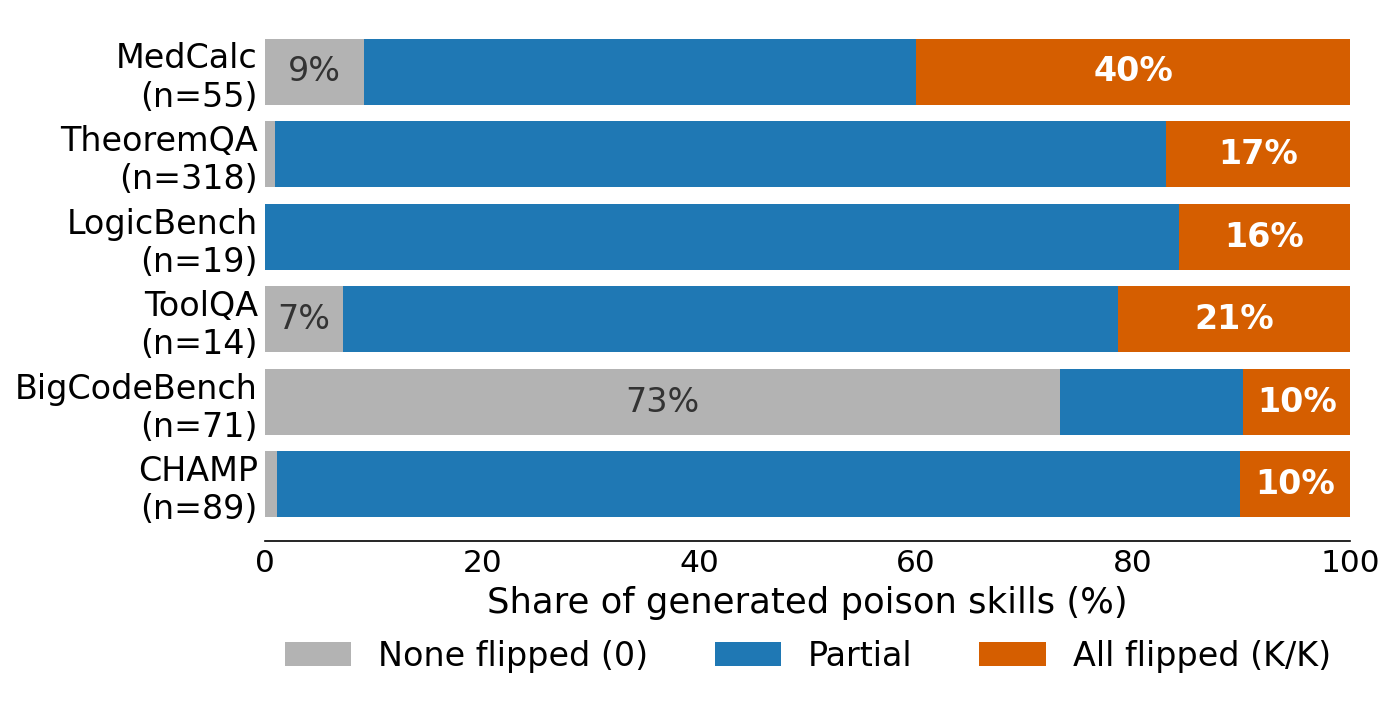}
    \caption{Distribution of generated poison skills by synthetic query flips. None flipped (grey) indicates that the poisoned skill changes none of the candidate answers; Partial (blue) indicates that it changes some but not all candidate answers; All flipped (orange) indicates that it changes all candidate answers.}
    \label{fig:construction-outcomes}
\end{figure}

\begin{table}[t]
\centering
\setlength{\tabcolsep}{3.5pt}
\renewcommand{\arraystretch}{1.08}
\caption{Oracle execution ASR (\%) of frozen \proceduralpoisoning payloads.
Qwen3-Coder-30B is the construction victim; cross-model
evaluation uses Llama-3.3-70B, Mistral-Small-24B, and
GPT-OSS-20B. Micro pools all pairs, while Macro averages
the six subsets equally.}
\label{tab:rq1-oracle-execution}
\begin{tabular}{@{}lrrrrr@{}}
\toprule
& & \textbf{Construct.} &
\multicolumn{3}{c}{\textbf{Cross-Model ASR}} \\
\cmidrule(lr){3-3}\cmidrule(l){4-6}
\textbf{Subset} & \textbf{Pairs}
& \textbf{Qwen3} & \textbf{Llama}
& \textbf{Mistral} & \textbf{GPT-OSS} \\
\midrule
TheoremQA  & 743  & 40.1 & 41.7 & 48.4 & 37.0 \\
LogicBench & 760  & 33.3 & 31.4 & 30.0 & 30.4 \\
ToolQA     & 1430 & 26.8 & 26.3 & 30.9 & 26.2 \\
MedCalc    & 1100 & 62.2 & 54.1 & 45.5 & 65.4 \\
CHAMP      & 382  & 13.1 & 17.5 & 19.5 & 6.8  \\
BigCode    & 1570 & 7.5  & 6.4  & 9.0  & 2.6  \\
\midrule
\textbf{Micro avg.} & 5985 & 32.9 & 31.0 & 31.3 & 31.9 \\
\textbf{Macro avg.} & 5985 & 30.5 & 29.6 & 30.5 & 28.1 \\
\bottomrule
\end{tabular}
\end{table}

We first assess whether \proceduralpoisoning payloads can alter skill outputs using only synthetic queries and pseudo-references. Construction is considered successful if a valid payload changes at least one synthetic-query answer. As shown in Figure~\ref{fig:construction-outcomes}, 89\% of the 566 generated payloads achieve at least one flip, with 72\% flipping some queries and 17\% flipping all queries. Effectiveness varies across tasks: 40\% of MedCalc payloads flip all queries, whereas 73\% of BigCode payloads flip none. These results demonstrate that \proceduralpoisoning payload construction is feasible for most target skills, but its effectiveness is task-dependent. For unsuccessful constructions, we retain a randomly selected valid candidate to evaluate potential effects on unseen queries.

\textit{Oracle execution on held-out queries.}
We next evaluate whether the frozen payloads induce incorrect task outcomes on the original SRA-Bench evaluation queries, which are not used during construction. Under oracle skill delivery, Table~\ref{tab:rq1-oracle-execution} shows that \proceduralpoisoning achieves a micro-averaged ASR of 32.9\% on the surrogate execution agent, Qwen3-Coder-30B. MedCalc has the highest ASR (62.2\%), while BigCode has the lowest (7.5\%). Unlike synthetic-query flips, these results measure incorrect answers against held-out ground truth, providing direct evidence of attack effectiveness beyond construction queries.

\textit{Cross-model transferability.}
The frozen payloads achieve micro-averaged ASRs of 31.0\%--31.9\% across Llama-3.3-70B, Mistral-Small-24B, and GPT-OSS-20B, close to the surrogate's 32.9\%. The task-level trends remain broadly consistent, with MedCalc generally more vulnerable and BigCode more resistant. This demonstrates that \proceduralpoisoning payloads transfer across execution models without target-model feedback.

Overall, \method can construct \proceduralpoisoning payloads for most eligible skills using only public skill content, synthetic queries, and pseudo-references. Oracle evaluation on the SRA-Bench user queries and ground truth shows that our generated poison skills can attack successfully across open-source models.

\textbf{RQ2: Can predicted-request metadata enable end-to-end attack success without retriever feedback?}

\label{sec:rq2}

\begin{table*}[t]
\caption{Full-registry retrieval and end-to-end attack effectiveness (\%) on
SRA-Bench with BGE-M3 as the first-stage retriever.
$\mathrm{ASR}_{\mathrm{Flip}}$ is the rate at which a poisoned skill outranks
its clean counterpart, while $\mathrm{E2E\text{-}ASR@1}$ is the
noise-corrected attack success rate under direct top-1 routing.}

\label{tab:rq2-retrieval-bgem3}
\begin{tabular}{llccccccc}
\toprule
& &
\multicolumn{3}{c}{\textbf{Retrieval}} &
\multicolumn{4}{c}{\textbf{E2E-ASR@1 (Top-1 Routing)}} \\
\cmidrule(lr){3-5}\cmidrule(lr){6-9}

\makecell[l]{\textbf{Subset}}
& \makecell[l]{\textbf{Method}}
& \makecell[c]{\textbf{PHit@1}}
& \makecell[c]{\textbf{PHit@20}}
& \makecell[c]{$\mathbf{ASR}_{\mathrm{Flip}}$}
& \makecell[c]{\textbf{Qwen3-Coder}\\\textbf{30B}}
& \makecell[c]{\textbf{Llama-3.3}\\\textbf{70B}}
& \makecell[c]{\textbf{Mistral-Small}\\\textbf{24B}}
& \makecell[c]{\textbf{GPT-OSS}\\\textbf{20B}}
\\

\midrule
\multirow{4}{*}{TheoremQA}
& No Opt.     & 35.4 & 87.1 & 49.8 & 11.3 & 10.9 & 13.3 & 12.3 \\
& HotFlip     & \textbf{54.8} & 89.5 & 71.5 & \textbf{16.7} & \textbf{16.8} & 19.6 & 17.8 \\
& GCG         & 53.8 & 89.6 & \textbf{72.1} & 16.2 & 16.5 & \textbf{19.9} & \textbf{17.9} \\
& \method & 49.3 & \textbf{91.5} & 67.0 & 14.4 & 14.6 & 17.9 & 16.6 \\
\midrule
\multirow{4}{*}{LogicBench}
& No Opt.     &  0.5 & 17.8 &  9.9 &  0.1 &  0.1 &  0.0 &  0.2 \\
& HotFlip     &  2.5 & 27.0 & 29.6 &  1.1 &  1.4 &  1.2 &  0.8 \\
& GCG         &  2.2 & 25.7 & 24.9 &  1.0 &  1.2 &  1.0 &  0.7 \\
& \method & \textbf{8.9} & \textbf{59.7} & \textbf{69.3} & \textbf{3.1} & \textbf{3.4} & \textbf{3.4} & \textbf{2.6} \\
\midrule
\multirow{4}{*}{ToolQA}
& No Opt.     &  2.1 &  8.3 & 11.6 &  0.2 &  0.1 &  0.2 &  0.2 \\
& HotFlip     &  4.3 & 16.9 & 20.5 &  0.7 &  0.5 &  0.8 &  0.6 \\
& GCG         &  3.2 & 15.7 & 18.0 &  0.5 &  0.4 &  0.6 &  0.5 \\
& \method & \textbf{13.1} & \textbf{25.8} & \textbf{33.1} & \textbf{1.7} & \textbf{1.5} & \textbf{2.9} & \textbf{1.1} \\
\midrule
\multirow{4}{*}{MedCalc}
& No Opt.     & 34.2 & 90.7 & 45.9 & 22.5 & 19.9 & 13.8 & 22.1 \\
& HotFlip     & 58.7 & 92.5 & 81.0 & 32.6 & 28.0 & 23.7 & 32.0 \\
& GCG         & \textbf{59.7} & 92.7 & \textbf{82.7} & \textbf{33.0} & \textbf{28.5} & \textbf{24.3} & \textbf{32.8} \\
& \method & 53.6 & \textbf{94.8} & 81.5 & 29.3 & 24.4 & 23.0 & 28.1 \\
\midrule
\multirow{4}{*}{CHAMP}
& No Opt.     &  9.4 & 51.1 & 38.6 &  1.2 &  1.3 &  1.2 &  1.5 \\
& HotFlip     & 16.1 & 56.5 & 61.0 &  4.3 &  4.5 &  3.9 &  2.4 \\
& GCG         & 17.9 & 56.5 & 62.8 &  4.0 &  4.5 &  3.7 &  2.3 \\
& \method & \textbf{20.2} & \textbf{65.9} & \textbf{76.7} & \textbf{4.8} & \textbf{5.2} & \textbf{4.9} & \textbf{3.0} \\
\midrule
\multirow{4}{*}{BigCode}
& No Opt.     & 13.9 & 73.6 & 47.9 &  1.5 &  1.1 &  2.0 &  1.1 \\
& HotFlip     & 22.1 & 77.2 & 63.7 &  2.3 &  1.7 &  2.3 &  1.5 \\
& GCG         & 22.7 & 76.5 & 61.1 & \textbf{2.6} &  1.9 &  2.4 & \textbf{1.6} \\
& \method & \textbf{32.1} & \textbf{86.1} & \textbf{80.0} & \textbf{2.6} & \textbf{2.2} & \textbf{2.9} &  1.4 \\
\midrule
\multirow{4}{*}{\textbf{Overall}}
& No Opt.     & 15.8 & 52.3 & 31.9 &  6.7 &  6.0 &  5.3 &  6.7 \\
& HotFlip     & 26.4 & 57.5 & 51.6 & \textbf{10.2} & \textbf{9.1} &  8.8 &  9.9 \\
& GCG         & 26.4 & 57.0 & 50.3 & \textbf{10.2} & \textbf{9.1} &  8.8 & \textbf{10.0} \\
& \method & \textbf{30.0} & \textbf{67.6} & \textbf{64.0} &  9.8 &  8.7 & \textbf{9.4} &  9.3 \\
\bottomrule
\end{tabular}
\end{table*}

\begin{table}[t]
\centering
\caption{End-to-end attack effectiveness (\%) in Claude Code with Opus 4.8,
evaluated on 120 queries per dataset. PHit@$k$ (PoisonHit@$k$) is the rate at
which a poisoned skill appears among the top-$k$ retrieved results; Select is
the poison-selection rate, and ASR is the end-to-end attack success rate.}
\label{tab:e2e-asr-claude}

\setlength{\tabcolsep}{2.5pt}
\renewcommand{\arraystretch}{1.05}

\begin{tabular}{@{}lccccc@{}}
\toprule
& &
\multicolumn{2}{c}{\textbf{Retrieval}}
& \textbf{Selection}
& \textbf{End-to-end} \\
\cmidrule(lr){3-4}
\cmidrule(lr){5-5}
\cmidrule(lr){6-6}

\textbf{Dataset}
& \textbf{\#Skills}
& \textbf{PHit@1}
& \textbf{PHit@20}
& \textbf{Select}
& \textbf{ASR} \\
\midrule
MedCalc    & 55 & 50.8 & 95.0 & 77.5 & 38.3 \\
TheoremQA  & 98 & 51.7 & 93.3 & 63.3 &  8.3 \\
CHAMP      & 65 & 24.2 & 65.8 & 39.2 &  0.8 \\
LogicBench & 19 & 10.0 & 63.3 & 31.7 &  5.0 \\
BigCode    & 67 & 45.8 & 95.8 & 60.8 &  5.0 \\
ToolQA     & 14 & 12.5 & 25.8 & 24.2 &  3.3 \\
\midrule
\textbf{Overall}
& \textbf{318}
& \textbf{32.5}
& \textbf{73.2}
& \textbf{49.4}
& \textbf{10.1} \\
\bottomrule
\end{tabular}
\end{table}

We evaluate whether fixed \proceduralpoisoning payloads remain effective under full-registry competition. We consider two workflows: (1) executing the top-ranked skill retrieved by BGE-M3 and (2) allowing Opus 4.8 to select and execute a skill from the top-20 retrieved candidates.

\textit{Retrieval placement.}
Table~\ref{tab:rq2-retrieval-bgem3} shows that \method improves PoisonHit@1 from 15.8\% (No Opt.) to 30.0\%, while $\mathrm{ASR}_{\mathrm{Flip}}$ increases from 31.9\% to 64.0\%. It also outperforms HotFlip and GCG in overall retrieval placement, achieving a 3.6-point advantage in PoisonHit@1 and the highest PoisonHit@20 (67.6\%). These results demonstrate that predicted-request metadata improves poison placement without retriever feedback, although gradient-based baselines remain stronger on some task families.

\textit{End-to-end effectiveness with top-1 routing.}
Improved retrieval placement also translates into higher attack success. Across four execution models, \method achieves E2E-ASR@1 of 8.7\%--9.8\%, compared with 5.3\%--6.7\% for No Opt. Its performance is comparable to HotFlip and GCG, differing by at most 0.7 percentage points. Thus, feedback-free metadata augmentation increases actual behavioral compromise under full-registry retrieval, rather than merely improving retrieval rankings.

\textit{End-to-end effectiveness with agent selection.}
We further evaluate a complete retrieval--selection--execution workflow using Claude Code with Opus 4.8. For 720 sampled queries, BGE-M3 retrieves the top-20 candidates, from which Opus 4.8 independently selects and executes one skill. As shown in Table~\ref{tab:e2e-asr-claude}, poisoned skills appear in the top-20 for 73.2\% of queries and are selected for 49.4\%, ultimately achieving 10.1\% E2E-ASR. Effectiveness varies substantially across tasks: MedCalc reaches 38.3\% E2E-ASR, whereas BigCode reaches only 5.0\% despite a 60.8\% poison-selection rate. This contrast shows that successful skill selection does not necessarily lead to behavioral compromise.

Overall, \method increases poison placement and end-to-end attack success without retriever feedback. Agent-controlled selection and execution constrain attack effectiveness but do not eliminate the threat.

\begin{table*}[t]
\centering
\small
\setlength{\tabcolsep}{4pt}
\caption{Poisoned-skill placement (PoisonHit@$1$, \%) before and after
reranking the top-20 BGE-M3 candidates.}
\label{tab:rerank20-bgem3-pseudo}
\resizebox{\textwidth}{!}{%
\begin{tabular}{l c cccc ccc}
\toprule
& &
\multicolumn{4}{c}{\textbf{Cross-encoder Reranking (Pointwise)}} &
\multicolumn{3}{c}{\textbf{LLM Listwise Reranking}} \\
\cmidrule(lr){3-6}\cmidrule(lr){7-9}
\textbf{Subset}
& \textbf{Retrieval}
& \textbf{SR-Rerank-0.6B}
& \textbf{BGE-Reranker-Large}
& \textbf{BGE-Reranker-v2-M3}
& \textbf{Jina-Reranker-v2}
& \textbf{Qwen3-4B}
& \textbf{Qwen3-14B}
& \textbf{Qwen3-32B} \\
\midrule
MedCalc
& 53.6 & 32.7 & 11.5 & 12.0 & 14.5 & 65.3 & 59.9 & 63.0 \\
TheoremQA
& 49.3 & 66.9 & 62.7 & 63.5 & 66.8 & 54.0 & 54.8 & 51.0 \\
CHAMP
& 20.2 & 48.4 & 19.7 & 35.0 & 41.7 & 32.7 & 43.9 & 39.9 \\
BigCode
& 32.1 & 49.3 & 21.4 & 11.5 & 50.2 & 26.4 & 36.1 & 41.5 \\
LogicBench
&  8.9 & 15.9 & 15.8 & 16.3 &  9.3 & 13.6 & 26.8 & 28.7 \\
ToolQA
& 13.1 & 21.0 & 22.1 & 23.1 & 21.0 & 23.1 & 22.1 & 23.6 \\
\midrule
\textbf{Overall}
& \textbf{30.0}
& \textbf{35.7}
& \textbf{24.5}
& \textbf{23.8}
& \textbf{30.8}
& \textbf{36.0}
& \textbf{38.9}
& \textbf{40.6} \\
\bottomrule
\end{tabular}%
}
\end{table*}

\textbf{RQ3: Do these placement gains transfer across retrieval and skill-selection models?}

We evaluate the transferability of \method across different retrievers and skill selectors, keeping the poisoned payloads and retrieval metadata fixed across all configurations.

\textit{Transferability across retrievers.}
Table~\ref{tab:method-victim} compares PoisonHit@$1$ across dense and sparse retrievers. \method outperforms HotFlip and GCG on all evaluated dense retrievers, including the held-out models, demonstrating that its placement gains are not specific to the surrogate embedding model. Among sparse retrievers, \method also performs best on BM25 ($31.5\,\%$), but falls below both baselines on SPLADE-v3 ($15.5\,\%$). This contrast suggests that transferability depends on the retrieval architecture rather than simply whether it uses dense or sparse representations.

\textit{Effect of skill selection.}
Table~\ref{tab:rerank20-bgem3-pseudo} evaluates seven selectors that rerank the top-20 candidates retrieved by BGE-M3. Compared with the initial PoisonHit@$1$ of $30.0\,\%$, post-selection PoisonHit@$1$ ranges from $23.8\,\%$ to $40.6\,\%$, with five of the seven selectors increasing poisoned-skill placement. Within the Qwen3 listwise family, overall PoisonHit@$1$ increases with model size, from $36.0\,\%$ (4B) to $40.6\,\%$ (32B). However, the effects vary across datasets and selectors. For example, on MedCalc, pointwise rerankers reduce poison placement, whereas all three listwise selectors increase it. These results indicate that downstream selection can either suppress or amplify retrieval poisoning, and stronger selectors do not necessarily provide greater robustness.

Overall, \method transfers across most evaluated retrievers, while downstream skill selection does not consistently mitigate poisoned-skill placement.

\begin{table}[htbp]
\centering
\setlength{\tabcolsep}{5pt}
\caption{PoisonHit@1 (\%) across retrievers on SRA-Bench. HotFlip and GCG are optimized against the BGE-large-en-v1.5 white-box surrogate ($\dagger$), whereas \method uses pseudo-query metadata augmentation without retriever feedback; all other retrievers are held out.}
\label{tab:method-victim}

\resizebox{\columnwidth}{!}{%
\begin{tabular}{llccc}
\toprule
\textbf{Retriever} & \textbf{Type}
& \textbf{HotFlip} & \textbf{GCG} & \textbf{\method} \\
\midrule
BGE-large-en-v1.5$^\dagger$ & Dense          & 24.6 & 25.3 & \textbf{28.4} \\
SR-Embed-0.6B  & Dense          & 23.9 & 24.4 & \textbf{30.2} \\
BGE-M3                      & Dense          & 26.4 & 26.4 & \textbf{30.0} \\
Contriever          & Dense          & 14.6 & 14.2 & \textbf{20.4} \\
BM25                        & LSR & 23.2 & 23.4 & \textbf{31.5} \\
SPLADE-v3                   & LSR & 17.8 & 18.4 & \textbf{15.5} \\
\bottomrule
\end{tabular}%
}
\end{table}

\textbf{RQ4: Can existing defenses detect procedural poisoning as effectively as overt malicious skills?}

\begin{figure}
    \includegraphics[width=1\linewidth]{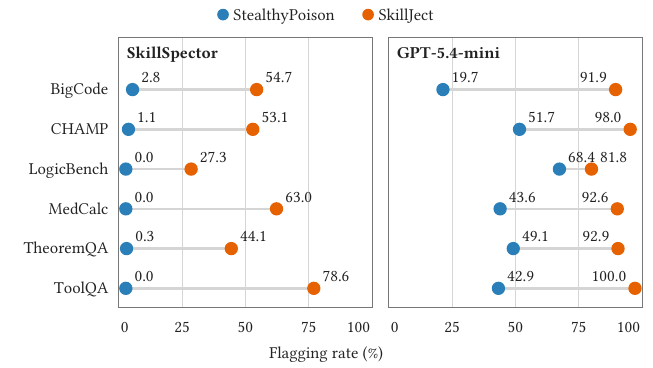}
    \vspace{2pt}
    {\footnotesize
    \textit{Other Defense Methods.} Cisco Skill Scanner yields overall flagging rates of 0.4\%, 0.4\%,
    and 0.2\% for clean, \method, and SkillJect packages, respectively. VirusTotal flags no packages in any subset.
    \par}


    \caption{Defense evaluation of \method and SkillJect across six benchmark subsets. Flagging rate denotes the percentage of skills flagged as malicious by each detector; full results, including clean skills and additional detectors, are reported in Appendix~\ref{app:defense_evaluation}.}
    \label{fig:defence_evaluation}
\end{figure}

We evaluate four representative pre-execution defenses: SkillSpector~\cite{pazskillspector}, Cisco Skill Scanner~\cite{cisco2026skillscanner}, VirusTotal file scanning~\cite{virustotalHowItWorks}, and GPT-5.4-mini~\cite{openai2026gpt54mini} as an LLM-based semantic filter, to examine whether they can identify poisoned and malicious skills before execution. The false-positive rate (FPR) on clean
skills and the detection rate (DetR) on poisoned skills are reported in Figure~\ref{fig:defence_evaluation}, which shows that all four defenses have
an overall FPR below $1\%$ on clean reference skills. The low but
nonzero FPR shows that a small number of benign packages also trigger
the detectors. The defenses perform very differently on SkillJect
packages. These packages contain more explicit malicious behaviors.
GPT-5.4-mini detects $93.1\%$ of SkillJect packages, while SkillSpector
detects $50.0\%$. Both detectors perform much better on SkillJect than
on \method, indicating that explicit malicious behaviors are easier
to recognize than changes embedded in the task procedure. Cisco Skill
Scanner and VirusTotal remain near zero. This result is consistent with
the packages rarely triggering their configured rules or malware-engine
signatures.

\method is substantially harder to detect. SkillSpector, Cisco Skill
Scanner, and VirusTotal detect only $0.7\%$, $0.4\%$, and $0.0\%$,
respectively. GPT-5.4-mini performs better, with a DetR of $45.8\%$, but
still misses a majority of the poisoned packages. It more often detects
explicit contradictions, reversed rules, fixed-answer cues, and
instructions to skip validation. It is less reliable when a modified
formula, constant, branch, or interpretation rule resembles a plausible
task procedure. We count a missed case as error-aware when the judge's
reason explicitly identifies an error, inconsistency, incorrect formula,
or implementation bug. By this definition, $59$ of the $307$ missed
cases are error-aware, yet the judge still labels them benign because
malicious intent is unclear. Thus, package-level semantic
screening can recognize visible signs of sabotage, but it cannot
reliably distinguish subtle poisoning from an ordinary implementation
error without a trusted behavioral reference.

These results show that pre-execution scanning can reduce the risk from
overt malicious skills, but it does not close the \proceduralpoisoning attack surface.
Defending against \proceduralpoisoning requires behavioral validation, such as testing
candidate skills on reference requests or comparing their outputs with
trusted implementations, rather than relying on package inspection
alone.

Existing pre-execution defenses can detect malicious skills but often fail to identify procedurally poisoned skills. Defending against \proceduralpoisoning therefore requires scanners that can verify the correctness of decision-critical procedural rules, rather than merely inspect malicious content or artifacts.

\section{Related Work}
\label{sec:related-work}
\textit{Retrieval poisoning and indirect prompt injection.} Retrieval poisoning attacks insert or optimize passages so that they rank highly for target queries~\cite{zhong2023adversarial,DBLP:conf/acl/SuNC25,DBLP:conf/naacl/WangWCH25}. Their optimization methods include HotFlip-style token substitutions~\cite{ebrahimi2018hotflip}. PoisonedRAG uses retrieved adversarial evidence to steer generated answers, while AgentPoison targets an agent's memory or RAG knowledge base; more recent work optimizes injected content for natural user queries~\cite{zou2024poisonedrag,DBLP:conf/nips/ChenXXSL24,DBLP:journals/corr/abs-2601-07072}. Indirect prompt injection instead places malicious instructions in external content such as webpages, documents, or tool outputs and can cause agents to take unintended actions~\cite{greshake2023injection,liu2024promptinjection,debenedetti2024agentdojo,zhan2024injecagent,zhang2025agentsecuritybench}. These attacks manipulate evidence, memory, or external content read by the agent. We study executable skills that compete against a full benign registry and alter the task procedure after selection.

\textit{Tool retrieval and registry manipulation.} Large tool use benchmarks study API discovery and invocation~\cite{li2023apibank,patil2023gorilla,qin2023toolllm}, while ToolRet, SkillsBench, and SRA-Bench study retrieval from large tool or skill collections~\cite{shi2025toolret,li2026skillsbench,su2026skill}. ToolHijacker injects a malicious tool document into this retrieval-and-selection pipeline to steer the agent toward an attacker-controlled tool~\cite{shi2025toolhijacker}. The Attractive Metadata Attack (AMA) optimizes tool names, descriptions, and parameter schemas to make a malicious tool more likely to be selected~\cite{DBLP:conf/nips/MoHLL25}. Recent work on malicious tools and MCP servers also studies poisoned tool descriptions, unsafe implementations, and hidden behaviors in tools already exposed to an agent~\cite{wang2026mcptox,hu2026maltool}. Other work shows that textual metadata in skill registries can influence discovery, selection, and governance decisions~\cite{saha2026underhood}. These studies establish that both tool metadata and tool implementations can be adversarial. ToolHijacker and AMA manipulate metadata to steer agents toward malicious tools. We evaluate poisoned skills from retrieval to execution and consider twin and parasitic placement, along with \proceduralpoisoning that changes the task itself instead of adding a separate malicious action.

\textit{Poisoned skill attacks.} Recent studies show that agent skills can contain malicious instructions. A large-scale study of public registries also finds malicious skills that combine adversarial documentation with executable helper scripts~\cite{liu2026not}. Skill-Inject evaluates whether agents follow obvious or context-dependent malicious instructions embedded in skill files, including variants in which an attacker modifies the skill body, adds an auxiliary script, or controls the skill description~\cite{schmotz2026skillinject}. SkillJect automates poisoned-skill construction by hiding an operational payload in auxiliary artifacts and iteratively refining the instructions that induce the agent to execute it~\cite{jia2026skillject}. Prior studies assume that the poisoned skill has already been loaded or selected. We additionally ask whether it can be retrieved from the registry, selected, and executed. Procedural poisoning keeps the skill's interface and artifacts unchanged but alters how the task is performed instead of adding a separate malicious action.

\textit{Pre-execution defenses for agent tools and skills.}
Prior work proposes defenses for agentic prompt injection, including 
instruction hierarchy, control-flow isolation, guard agents, and
LLM-as-a-judge filters~\cite{wallace2024instructionhierarchy,deb2025camel,xiang2024guardagent,gu2024llmasjudge}. These defenses aim to prevent
untrusted instructions or unsafe tool behavior from influencing an agent at
runtime. StruQ separates trusted instructions from untrusted data~\cite{DBLP:conf/uss/ChenPS025}, while TRUSTDESC derives tool descriptions from implementations and verifies their behavioral claims~\cite{DBLP:journals/corr/abs-2604-07536}. These methods reduce prompt or metadata manipulation, but they do not directly check whether a skill preserves the intended task procedure. Static and behavioral scanners provide a complementary
pre-execution layer by flagging suspicious code, metadata, or capability
use before a tool or skill is invoked. We evaluate these defenses on poisoned
skill packages. Unlike overt policy violations, \proceduralpoisoning can preserve the public interface while changing the task procedure, so detecting it requires
behavioral validation against trusted reference behavior.

\section{Discussion}
\label{sec:discussion}

Our findings highlight two security implications for skill-augmented agents.

\textbf{Retrieval Relevance Does Not Imply Trustworthiness.}
Retrievers and skill selectors prioritize task relevance rather than procedural integrity. Our results show that poisoned metadata can improve skill placement and that downstream reranking may further amplify this effect. Therefore, retrieval and selection cannot be treated as implicit security defenses. Moreover, neither retrieval success nor oracle execution success alone establishes end-to-end risk: a successful attack requires the poisoned skill to be selected and its corrupted procedure to affect the final answer.

\textbf{Successful Execution Does Not Guarantee Procedural Correctness.}
Procedural poisoning demonstrates that a skill can preserve its advertised task, interface, and executable artifacts while producing incorrect results through corrupted decision rules. This exposes a limitation of defenses focused on malicious instructions or unauthorized actions. Protecting against such attacks requires independent verification of skill-defined procedures, for example, against trusted formulas, reference implementations, or test cases. Such verification is particularly challenging when no authoritative reference is available.

\section{Conclusion}
In this paper, we presented \method, an end-to-end attack against
skill-augmented agents that targets both skill retrieval and procedural
execution. Our evaluation shows that procedural poisoning can induce
incorrect task outcomes while remaining difficult to detect because the
manipulation is embedded in otherwise legitimate task procedures. When
real user requests are unavailable, predicted-request metadata improves
poisoned-skill placement without retriever gradients or ranking feedback,
outperforming HotFlip and GCG across most evaluated retrievers. These
findings show that malicious-content detection and relevance-based routing
alone cannot secure public skill ecosystems. Effective defenses must
protect both registry routing and procedural integrity through stronger
metadata governance, retrieval-manipulation detection, and independent
validation of decision-critical rules.





\bibliographystyle{ACM-Reference-Format}
\bibliography{reference}
\appendix
\clearpage
\section{Open Science}
All attack and defense code, datasets, as well as LLM prompts, are available at \url{https://anonymous.4open.science/r/skillpoison-BC93}

\section{Ethical Considerations}
\label{app:ethical_considerations}
This work studies attacks on agent skills. It can help defenders find and
fix weaknesses in the retrieve--select--execute pipeline, but attackers
could also use these methods to create more effective poisoned skills.
Potentially affected stakeholders include agent users, skill developers,
registry operators, platform providers, and security researchers.

\textit{Controlled, safe evaluation.}
All experiments use offline benchmarks and locally constructed test
registries. We do not submit poisoned skills to public marketplaces,
target real users, or attack deployed systems. The skills and requests
come from public benchmarks or are generated for research. Our payloads
operate only on synthetic or decoy data. They do not collect real
credentials, personal data, private user content, or proprietary
information, and they do not send sensitive data to outside servers.
Thus, our experimental process does not cause harm to existing
infrastructures or users.

\textit{Defense evaluation and artifacts.}
For pre-execution defense evaluation, we submit only research skill
packages and remove API keys, credentials, private paths, and other
sensitive information before using external services. We release
artifacts needed for reproducibility, such as prompts, metadata, scripts,
and aggregate results, while excluding live credentials, private
endpoints, and ready-to-deploy attack packages.

\textit{Purpose and limitations.}
Our goal is to help improve defenses for skill registries. Our results
show that current filters may miss hidden changes to how a skill performs
a task. These failures could cause harm when agents are used for
technical, medical, financial, or operational tasks. Because all tests
are local, the measured attack success rates may differ from those in
real systems. Real systems may have stronger defenses, but they may also
give agents access to sensitive data or important actions.

\section{Model Configurations and Access Assumptions}
\label{app:model-configurations}

\textbf{Access Assumptions.}
We distinguish white-box and black-box access separately for execution
and retrieval. For execution, white-box access provides the surrogate
agent's outputs and execution traces during payload construction, whereas
black-box access provides no execution feedback for generating, refining,
or selecting payloads. For retrieval, white-box access permits access to
model parameters and gradients, whereas black-box access provides neither
model parameters nor retrieval scores, rankings, or feedback during
attack construction.

\textbf{Execution Models.}
Mistral-Small-3.2-24B-Instruct-2506~\cite{mistral2025small32} serves as
the attack model for generating candidate \proceduralpoisoning payloads.
Qwen3-Coder-30B~\cite{DBLP:journals/corr/abs-2505-09388} serves as the
surrogate execution agent for payload validation and refinement. After
construction, the selected payloads are frozen and evaluated on
Qwen3-Coder-30B as the construction-time reference, followed by
Llama-3.3-70B~\cite{DBLP:journals/corr/abs-2407-21783},
Mistral-Small-3.2-24B-Instruct-2506,
GPT-OSS-20B~\cite{agarwal2025gpt}, and
Opus 4.8~\cite{anthropic2026opus48} via
Claude Code~\cite{anthropic2026claudecode}. No outputs or execution
traces from these target evaluations are used for payload refinement.
Opus 4.8 is also used to generate plausible requests for retrieval
optimization.

\textbf{Retrieval Models.}
BGE-large-en-v1.5~\cite{DBLP:journals/corr/abs-2309-07597} serves as the
white-box surrogate retriever, while
BGE-M3~\cite{DBLP:journals/corr/abs-2402-03216} is the primary black-box
victim retriever. Additional victim retrievers include
SkillRouter-Embedding-0.6B~\cite{DBLP:journals/corr/abs-2603-22455},
Contriever-MSMARCO~\cite{DBLP:journals/tmlr/IzacardCHRBJG22},
BM25~\cite{DBLP:journals/ftir/RobertsonZ09}, and
SPLADE-v3~\cite{DBLP:journals/corr/abs-2403-06789}.

\textbf{Skill Selection Models.}
We rerank the top 20 candidates retrieved by BGE-M3 using seven models:
SkillRouter-Reranker-0.6B~\cite{DBLP:journals/corr/abs-2603-22455},
BGE-Reranker-Large~\cite{baai2023bgererankerlarge},
BGE-Reranker-v2-M3~\cite{li2023making,DBLP:journals/corr/abs-2402-03216},
Jina-Reranker-v2~\cite{jina2024rerankerv2}, and Qwen3-4B, Qwen3-14B,
and Qwen3-32B~\cite{DBLP:journals/corr/abs-2505-09388}.

\section{Defense Evaluation}
\label{app:defense_evaluation}

\begin{table}[H]
\centering
\small
\caption{Detection rates (\%) for clean, \method (SP), and SkillJect
(SJ) skills.}
\label{tab:detector_comparison}
\setlength{\tabcolsep}{9pt}
\renewcommand{\arraystretch}{1.25}

\begin{tabular}{@{}llrrr@{}}
\toprule
\textbf{Subset}
& \textbf{Detector}
& \textbf{Clean}
& \textbf{SP}
& \textbf{SJ} \\
\midrule

BigCode
& SkillSpector & 2.8 & 2.8 & 54.7 \\
& Cisco        & 2.8 & 2.8 & 1.2 \\
& GPT-5.4-mini & 1.4 & 19.7 & 91.9 \\
& VirusTotal   & 0.0 & 0.0 & 0.0 \\
\addlinespace[3pt]

CHAMP
& SkillSpector & 1.1 & 1.1 & 53.1 \\
& Cisco        & 0.0 & 0.0 & 0.0 \\
& GPT-5.4-mini & 0.0 & 51.7 & 98.0 \\
& VirusTotal   & 0.0 & 0.0 & 0.0 \\
\addlinespace[3pt]

LogicBench
& SkillSpector & 5.3 & 0.0 & 27.3 \\
& Cisco        & 0.0 & 0.0 & 0.0 \\
& GPT-5.4-mini & 0.0 & 68.4 & 81.8 \\
& VirusTotal   & 0.0 & 0.0 & 0.0 \\
\addlinespace[3pt]

MedCalc
& SkillSpector & 0.0 & 0.0 & 63.0 \\
& Cisco        & 0.0 & 0.0 & 0.0 \\
& GPT-5.4-mini & 0.0 & 43.6 & 92.6 \\
& VirusTotal   & 0.0 & 0.0 & 0.0 \\
\addlinespace[3pt]

TheoremQA
& SkillSpector & 0.3 & 0.3 & 44.1 \\
& Cisco        & 0.0 & 0.0 & 0.0 \\
& GPT-5.4-mini & 0.6 & 49.1 & 92.9 \\
& VirusTotal   & 0.0 & 0.0 & 0.0 \\
\addlinespace[3pt]

ToolQA
& SkillSpector & 0.0 & 0.0 & 78.6 \\
& Cisco        & 0.0 & 0.0 & 0.0 \\
& GPT-5.4-mini & 0.0 & 42.9 & 100.0 \\
& VirusTotal   & 0.0 & 0.0 & 0.0 \\
\midrule

\textbf{Overall}
& SkillSpector & 0.9 & 0.7 & 50.0 \\
& Cisco        & 0.4 & 0.4 & 0.2 \\
& GPT-5.4-mini & 0.5 & 45.8 & 93.1 \\
& VirusTotal   & 0.0 & 0.0 & 0.0 \\
\bottomrule
\end{tabular}
\end{table}

\section{Retrieval Baseline Configurations}
\label{app:retrieval-baselines}

\begin{table*}[t]
\centering
\caption{
Overview of the six source datasets used to construct SRA-Bench.
Raw Queries and Raw Skills denote the original benchmark statistics.
Raw Gold Skills refers only to the task-relevant reference skills derived
from the six source datasets and excludes the 25,626 distractor skills in
the retrieval registry. Eval.\ Queries and Eligible Skills denote the
instances retained after validation. Eval.\ Pairs counts the resulting
query--skill pairs; this number may exceed Eval.\ Queries for datasets
with multiple gold skills per query.
}
\label{tab:source-datasets}
\small
\setlength{\tabcolsep}{4.5pt}

\begin{tabular}{@{}llrrrrcl@{}}
\toprule
\textbf{Dataset}
& \textbf{Capability Type}
& \textbf{Raw Queries}
& \textbf{Eval.\ Queries}
& \textbf{Raw Gold Skills}
& \textbf{Eligible Skills}
& \textbf{Mapping}
& \textbf{Evaluation} \\
\midrule
\textsc{TheoremQA}
& Theorem Application
& 747 & 743 & 320 & 318 & Single & Rule-Based \\
\textsc{LogicBench}
& Logical Reasoning
& 760 & 760 & 19 & 19 & Single & Rule-Based \\

\textsc{ToolQA}
& Tool-Use Workflows
& 1,430 & 1,430 & 14 & 14 & Single & Rule-Based \\

\textsc{MedCalc}
& Medical Calculators
& 1,100 & 1,100 & 55 & 55 & Single & Rule-Based \\

\textsc{CHAMP}
& Mathematical Concepts
& 223 & 223 & 89 & 89 & Multi & Rule-Based \\

\textsc{BigCode}
& Software Libraries
& 1,140 & 988 & 139 & 71 & Multi & Execution \\
\midrule

\textbf{Total}
& --
& \textbf{5,400}
& \textbf{5,244}
& \textbf{636}
& \textbf{566}
& --
& -- \\
\bottomrule
\end{tabular}
\end{table*}

We evaluate three retrieval baselines against \method, keeping the
validated \proceduralpoisoning payloads and executable artifacts fixed
across all methods. For each skill, we generate 20 plausible requests
from its public content without access to evaluation queries. \method
directly inserts these requests into its retrieval metadata, while
HotFlip~\cite{ebrahimi2018hotflip} and
GCG~\cite{zou2023universal} use them as optimization queries. The
\emph{No Opt.} baseline preserves the original retrieval metadata.

Following prior work on retrieval-corpus
poisoning~\cite{zhong2023adversarial}, HotFlip and GCG optimize the
retrieval keyword field by maximizing the mean cosine similarity between
the poisoned skill's BGE-large-en-v1.5 embedding and the embeddings of
the plausible requests. Both run for 60 optimization steps. HotFlip
searches the full vocabulary, whereas GCG evaluates 64 gradient-selected
candidates per iteration. In contrast, \method requires no retrieval
scores, gradients, or feedback from either the surrogate or victim
retriever. This comparison isolates the effectiveness of feedback-free
metadata construction.


\end{document}